\documentclass[conference]{IEEEtran}

\usepackage{url}
\usepackage{booktabs}
\usepackage{rotating}
\usepackage{array}
\usepackage{ragged2e}
\usepackage{comment}
\usepackage{pdflscape}
\usepackage{cite}

\usepackage{amsmath,amssymb,amsfonts}
\usepackage{algorithmic}
\usepackage{graphicx}
\usepackage{textcomp}
\usepackage{xcolor}
\def\BibTeX{{\rm B\kern-.05em{\sc i\kern-.025em b}\kern-.08em
    T\kern-.1667em\lower.7ex\hbox{E}\kern-.125emX}}
\begin{document}

\title{Wireguard: An Efficient Solution for Securing IoT Device Connectivity\\
\footnotesize  Regular Research Paper (CSCI-RTMC)}

\author{\IEEEauthorblockN{1\textsuperscript{st} Haseebullah Jumakhan}
\IEEEauthorblockA{\textit{Independent Researcher} \\
\textit{Not Affiliated}\\
Ajman, United Arab Emirates \\
0009-0005-5461-5531}
\and
\IEEEauthorblockN{2\textsuperscript{nd} Amir Mirzaeinia}
\IEEEauthorblockA{\textit{Computer Science and Engineering Department} \\
\textit{University of Northern Texas}\\
Denton, Texas, United States \\
amir.mirzaeinia@unt.edu}
}

\maketitle

\begin{abstract}

The proliferation of vulnerable Internet-of-Things (IoT) devices has enabled large-scale cyberattacks. Solutions like Hestia and HomeSnitch have failed to comprehensively address IoT security needs. This research evaluates if Wireguard, an emerging VPN protocol, can provide efficient security tailored for resource-constrained IoT systems. We compared Wireguard’s performance against standard protocols OpenVPN and IPsec in a simulated IoT environment. Metrics measured included throughput, latency, and jitter during file transfers. Initial results reveal Wireguard's potential as a lightweight yet robust IoT security solution despite disadvantages for Wireguard in our experimental environment. With further testing, Wireguard’s simplicity and low overhead could enable widespread VPN adoption to harden IoT devices against attacks. The protocol’s advantages in setup time, performance, and compatibility make it promising for integration especially on weak IoT processors and networks.
\end{abstract}

\begin{IEEEkeywords}
IoT security, Network security, VPN, VPN performance, Wireguard
\end{IEEEkeywords}

\section{Introduction}
With the general increase in smaller smart devices connected to the Internet, the IoT devices, there has also been a large increase in the risks inherent to their portability and connectivity. For example, in 2016, thousands of IoT devices were infected and then used as part of a botnet for a DDoS attack that targeted Twitter, Spotify, Paypal, and made unavailable other popular websites like CNN, New York Times, etc.~\cite{b1}. The Mirai botnet, which used default password lists to break into vulnerable IoT devices, is a great example of a large-scale attack that was discovered in 2016 \cite{b2}. More recently, the Verkada hack in March 2021 was another major event with serious complications for user privacy as a result of IoT security failure \cite{b1}. From the above real-world examples, the possibility of using millions of unprotected IoT devices for such malicious attacks, as mentioned above, is quite visible. There are also other types of attacks that are possible with such devices. A highly sensitive medical or military facility’s connected devices or smart thermostats could be hijacked to coerce people into what the hackers demand \cite{b3}.

We will discuss and then test the most popular secure tunneling solutions available that can be potentially used for securing IoT devices on the Internet. These include: 
\begin{enumerate}
    \item Widely used general-purpose security solutions such as Secure Socket Layer/Transport Layer Security (SSL/TLS) protocol that works at the application layer of the OSI Model to provide secure browsing on almost all web pages.
    \item Virtual Private Networks that work at the network or transport layer and also provide an encrypted tunnel, albeit with properties different from those of the SSL/TLS protocol.
\end{enumerate}
A detailed comparison of VPN and SSL/TLS is to follow after discussing two other solutions that have been proposed to solve the problem that this paper is discussing and are solutions too specific to IoT, which render them incomparable to SSL/TLS and VPNs. Namely, Hestia \cite{b4} and HomeSnitch \cite{b5} are two existing solutions highly specific to IoT devices. However, they have inherent problems of being too complex and providing limited security, and we will describe their properties in the next section and then compare the more general-purpose solutions, SSL/TLS and VPN, which are still viable options for IoT devices.

\section{Related Work and Background}

\subsection{Hestia}

Hestia tries to prevent the hijacking of devices on a local network by only allowing controller devices like a Google Home module, an Amazon Alexa, etc. to connect to controlled devices like cameras, door locks, smart lights, and such. Below are some issues with Hestia that render it unreliable from a security and ease of usability perspective: 

\begin{enumerate}
\item The controller devices can themselves have unreliable protection and can possibly be hijacked by an attacker. This happens when controllers are not updated and/or they have a vulnerability, which can be extremely difficult to cover since there are numerous manufacturers and they each have many more products. Since the controllers have access to all devices on the network, it is still risky to rely on such a setting.
\item With blocking the controlled devices from the Internet completely, the administrator/user creates a dependency on the controller devices. That is, the user has to be able to control the controller devices to control the controlled devices at the premise. This could increase the risk of damage to property if and when the controllers are hijacked and the owner is at a remote location since he/she is shut off from all the controlled devices by design. 
\end{enumerate}

\subsection{HomeSnitch}

HomeSnitch is another proposed solution, and it tries to learn normal device behavior for every IoT at a premise over a certain period of normal use and then effectively firewalls packets that have not routinely been sent or received in hopes of preventing and busting the hacking of these devices. Although it is an excellent way to think about the security problems of IoT's, some issues come to mind with HomeSnitch and make it unfavorable to use:
\begin{enumerate}
\item It is too complicated for the average user and can be quite time-consuming for sysadmins too, which make it extremely difficult to maintain. The maintenance and too much specificity makes it difficult because one will have to let the system learn every device's behavior and also monitor anomalies. If some less frequent but essential packets are blocked and not allowed in time, it could malfunction the IoT devices and cause sensor blinding; that is, it would render useful signals from some devices with infrequent updates as anomalies when they do provide a genuine update. 
\item This method could increase the risk of successful attacks as it can block certain infrequent updates that will be treated by the system as anomalies. Effectively not allowing the IoT devices that do update to update and increasing the attack surface for attackers.
\end{enumerate}

Digressing back to SSL/TLS vs.~VPN, the solutions popular across platforms in the current age, we will first narrow down our pick in between TLS/SSL and VPN in this section. In Section~\ref{sec:idea}, we will further narrow down a specific VPN solution, Wireguard, among the VPN technologies that are available. 

\subsection{Virtual Private Networks (VPN) vs. SSL/TLS}

\textbf{VPN.} A Virtual Private Network is an extension of a network (usually secured by encryption) to a remote location that is accessible by an authenticated user. The phrase "extension of a network" shall be emphasized here, as this is the main defining feature of a VPN. It means that a VPN does not just let you connect to a remote location — there are many other ways to do that - but that it lets you access a complete remote network from a remote location. 

The use of VPNs was popularized by businesses initially, but it did not take long to become popular among privacy-conscious people for its privacy features, which will be discussed later in this sub-section in Table~\ref{tab:Table 1}.

\noindent\textbf{SSL/TLS.} SSL/TLS, on the other hand, is a security protocol that provides reliable security for individual connections between two hosts separately from all other connections. In other words, they are only for a one-to-one connection and do not include properties of a VPN such as extending a network to a remote location or a firewall.

Why choose a VPN solution instead of TLS/SSL? Table~\ref{tab:Table 1} summarizes the key differences in capabilities between VPN and SSL/TLS that demonstrate VPN's advantages in securing an entire network of IoT devices. VPN provides an encrypted tunnel to not just connect to a remote system but to extend an entire network to a remote location. This enables users to consolidate control of all IoT devices through a single encrypted VPN tunnel with one firewall on the VPN server, reducing points of vulnerability. VPN also allows anonymous remote access, reducing visibility to an ISP or bad actors. In contrast, TLS only facilitates a point-to-point encrypted connection between two devices. Popular VPN protocols include OpenVPN and IPsec. However, limitations exist in applying them effectively to IoT contexts, as will be discussed in Section~\ref{sec:idea}.

\begin{table}
\centering
\resizebox{\linewidth}{!}{%
\begin{tabular}{>{\hspace{0pt}}p{0.481\linewidth}>{\hspace{0pt}}p{0.46\linewidth}} 
\toprule
TLS & VPN \\
\midrule
Uses TCP only & has the option to use UDP, implying a potential speed advantage. \\
Layer 4 (Transport Layer) and above & Layer 3 (Network Layer). Part of the Kernel, so runs faster because of close proximity to the processor architecture. \\
ISP knows both parties. Reducing privacy when controlling devices remotely & Anonymous IP address. ISP knows only one party (less monitoring, tracking, and reconnaissance hacking) \\
TLS is not used by many IoT devices by default & Acts as an extra layer of protection over devices just in case they do not use TLS by default, i.e., provides security for devices that are not configured securely. \\
Provides only a single connection. That is, TLS only provides a point to point connection between two devices, and it is not ideal for accessing a full network of devices from another location. & Provides a secure network of selected devices. A user can consolidate outside connections to one VPN server and control multiple devices with only one firewall on the VPN server. \\
Needs to use certificates. & can use other methods of authentication. \\
\bottomrule
\end{tabular}
}
\caption{Comparing TLS and VPN}
\label{tab:Table 1}
\end{table}

\subsection{Wireguard}

Wireguard is a new VPN protocol launched in 2017 by Jason A. Donenfeld. Donenfeld’s goal was to improve on OpenVPN and IPsec \cite{b6}. This VPN has been welcomed for numerous reasons, listed as follows:
\begin{enumerate}
\item Wireguard Lives in the Kernel: This greatly improves performance over VPN for the userspace layer is not at all involved in the resource-intensive processes of the VPN, keeping the protocol power-efficient. This is also an excellent property to have for many battery-powered IoT devices.
\item Smaller Codebase Provides Potentially Better Security: Wireguard is just 3,800 lines of code and hence, is easily auditable. This makes it easier to find security flaws in the protocol and also reduces the attack surface for hackers.
\item  Simplicity – Version-controlled Properties: Compared to other popular VPN protocols like OpenVPN and IPsec, Wireguard is much easier to configure on a system and get it running. The encryption (ChaCha20), authentication (Poly1305), and hashing (BLAKE2s) are all predetermined, and Wireguard does not allow for using various solution options as OpenVPN does. It is almost as simple to set up as SSH is.
\item Better Performance: This is an outstanding aspect of Wireguard for this project. The substantially better performance of a modern, easily auditable, lightweight VPN is greatly promising for the security of IoT’s.
\item Wireguard works with mesh networks: You do not necessarily need a central server device in Wireguard. This typically reduces time in half  when two nodes want to communicate, which is immensely useful in the case of IoT devices that are interconnected and make some/all decisions independently through preset commands of AI.
\end{enumerate}

Wireguard's major advantages are performance-related, and hence, could be a potential solution to the IoT risks problem. In other words, it could be a viable security-enhancing VPN solution that can work with a typical IoT's limited resources. For example, limited processing power, limited memory, et cetera will not be as big a problem for Wireguard as they would be for other heavy VPN solutions like OpenVPN and IPsec. Wireguard’s source code is just 3,800 lines, a lot less than competing protocols OpenVPN's 650,000 lines and IPsec's 400,000 lines \cite{b7}.

\begin{figure}[h!]
\centering
\includegraphics[width=1.\columnwidth]{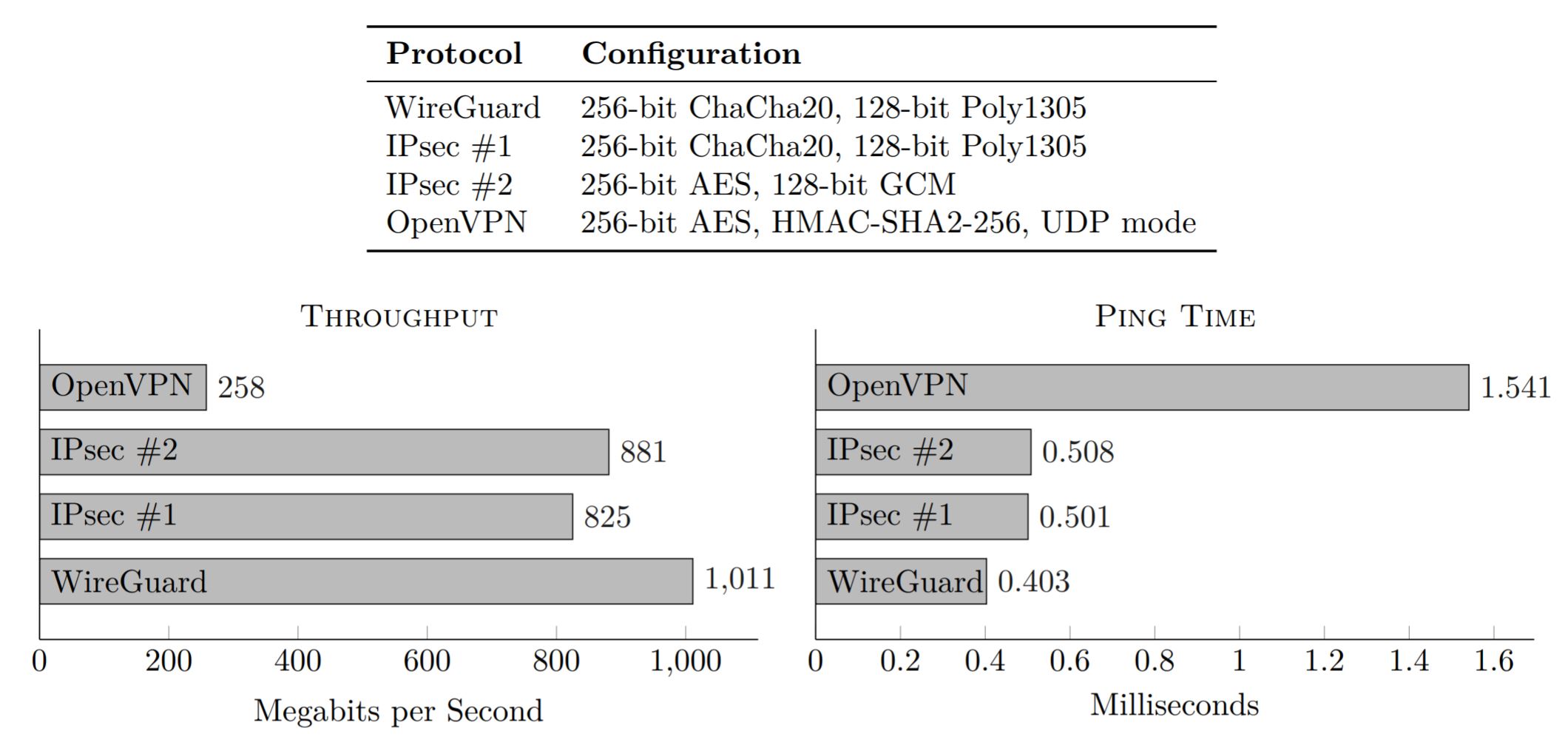}
\caption{Results of performance testing. Source: Wireguard whitepaper \cite{b6}}
\label{whitepaper} 
\end{figure}

Its simplicity is the main strength that enables it to be faster than other conventional protocols, which is visible in Figure ~\ref{whitepaper} sourced from Wireguard's whitepaper.

\subsection{Cryptographic Properties of Wireguard}

Wireguard adopts some of the latest cryptography solutions available, which give it some advantages:

\begin{enumerate}

\item Wireguard uses ChaCha20 for its encryption. ChaCha20 is the obsolete and vulnerable RC4’s replacement. It is a symmetric stream cipher that is quantum-resistant and uses only 64-bit keys. It is faster than most other ciphers. Smaller devices like the ones we are dealing with (IoT devices) can benefit from ChaCha20 because it is faster than AES in these devices, as these small devices do not have the AES-NI instructions that make it easier for the CPUs that have them to perform AES calculations. Wireguard does not need this configuration of the CPU architecture to be efficient, hence, it poses itself as a strong contender for these weak, and usually, cheap IoT devices \cite{b8} \cite{b9}. It can also be inferred that including many IoT devices, Wireguard stands a better chance of running efficiently on a set of random devices than protocols that use AES encryption.

\item Additionally, ChaCha is not vulnerable to cache-timing attacks \cite{b8} \cite{b9}. A cache-timing attack is when an attacker observing a cryptographic communication can decipher some information based on the time a process needs to process certain things. AES needs lookup tables to avoid this, and an unsophisticated implementation can make it vulnerable to cache-timing attacks, while ChaCha does not need this extra complexity to prevent cache-timing attacks.

\item In the case of ChaCha calculations, it is mostly just XORing and 4x4 matrix calculations. It keeps the calculations simple yet provides good diffusion and other necessary properties. It is being adopted by Google and others to use inside TLS and is being used exclusively on Wireguard \cite{b9}.

\item Wireguard uses Poly1305 for authentication. Poly1305 is a MAC. During the initial authentication, an increasing timestamp is used. A responder keeps the greatest timestamp of the peer even if a shared key is compromised. This helps against replay attacks \cite{b6}.

\item The routing that Wireguard is configured with has lists of approved peers, and all peers have the option of pre-sharing keys for quantum-proof encryption. Hence, the authentication in Wireguard is two-fold, both routing peers' IP lists and shared keys \cite{b6}.

\item Wireguard also gives the option of Passive Keepalive. This option can be changed on the fly by a peer without having to reconnect. When the responder has nothing to say and Passive Keepalive, every message that is received is replied with an encrypted but empty message of a certain length (according to the IPv4 and IPv6 headers). If they do not communicate over a long time, which is more than the specified timeout time, then they renegotiate \cite{b6}. This renegotiation is extremely quick and is much less chatty than the usual chatty keepalive algorithms found in other protocols.

\item Wireguard does not use certificates. An advantage of not using certificates is that it does not fall victim to renegotiation attacks. Wireguard can do this because of its preconfigured tables of trusted and authenticated IP addresses \cite{b6}.
\end{enumerate}

\section{Idea: Using Wireguard for IoT Security}
\label{sec:idea}

Now that VPN has been shown to have superior features in regards to securing a multitude of devices owned by a single entity and Wireguard has been introduced, it is paramount to show that Wireguard is an attractive solution for securing weak IoT devices that are likely to be vulnerable.

Conventional solutions like OpenVPN \cite{b10}, secure tunnels, etc.~have proven their reliability over time. In the case of OpenVPN, three properties of the said protocol make it extremely inefficient and cumbersome to implement on IoT devices. These are:

\begin{enumerate}

\item Complexity and High Overheads:  OpenVPN offers myriad configuration options, requiring multiple underlying software implementations for different setups. This demands more RAM and ROM than is optimal for lightweight IoT devices. As a result of the above complexity and the fact that OpenVPN has a large codebase, OpenVPN is not a straightforward protocol to work with and requires good performance devices to provide performance at full capacity.

\item Difficulty in Implementation and Maintenance: The complexity makes OpenVPN challenging to initially implement and maintain. OpenVPN connections tend to drop regularly \cite{b10}, and reconnecting the devices in the thousands could render it impractical for IoT's.

\item High Latency: The encryption and decryption in OpenVPN are done in the userspace. This results in additionally higher latencies \cite{b10}.
\end{enumerate}

IPsec, on the other hand, is another popular solution that is used in the form of a VPN in its tunneling mode \cite{b11}. However, it has its drawbacks if applied to IOT’s:
\begin{enumerate}
\item The most popular implementation of IPsec by Cisco needs a proprietary, closed-source framework to function.

\item Close Grouping of Devices on the Network: If you are connected to an IPsec network, it is possible to “gain access across the WAN to the corporate network.” So, a single infected device could pollute multiple devices on this network \cite{b12}. 
\item IPsec Traffic is Charged More in Some Places: Some ISP’s charge more IPsec traffic because it is usually attributed to business traffic \cite{b12}. \end{enumerate}

This is not to deny that OpenVPN and IPsec do have their strengths (high configurability, high compatibility with various platforms, and immense third-party security testing), but it is to show that OpenVPN is not a good solution to the issue of securing IoT devices.

\section{EXPERIMENTAL SETUP}

\subsection{Initial Setups}
To simulate the low-end capacity and power of a typical IoT device, we initially decided to use the third model of the Raspberry Pi, the Raspberry Pi 3B+, which has a four-core processor made of A53 chips at 1.4GHz, just one gigabyte of RAM, and a maximum throughput of 1 Gbits/s through Ethernet. The original speed comparison tests that are shown above from the Wireguard whitepaper were performed on powerful devices, but in order to show the advantages of Wireguard in the context of weak IoT devices, we planned to test the various VPN solutions on four Raspberry Pi devices more appropriate to the IoT context of this paper.

These four devices were set up using a Linux fork called Ubuntu Mate \cite{b13}, which is a fairly stable version of Linux that works on the Raspberry Pi. The testing will be done for the three separate protocols — Wireguard, OpenVPN, and IPsec — in turns to avoid any competition for the available bandwidth. Please note that the open-source version of IPsec, OpenConnect, will be used for IPsec. This is due to the licensing issues with IPsec discussed in Section III. This plan was majorly changed because of the physical limitations faced at the premises of the research. These limitations and the workaround are discussed below.

The VPN testing setup aimed to test all protocols under identical conditions and hardware. However, a challenge emerged. The available ISP actively throttled and blocked certain VPN protocols that bypassed restrictions. OpenVPN, one of the protocols tested, fell into the throttled category. OpenConnect was also throttled, but less than OpenVPN. Surprisingly, Wireguard did not seem affected by throttling despite being UDP-only. This might have happened because the ISP does not recognize it as a major tool yet and has not actively blocked it. Upon testing with iPerf, \textit{Wireguard yielded data transfer performance at full capacity of the available link}, but OpenVPN achieved only less than 1 percent of the capacity of the link while OpenConnect stayed somewhere in between. This data does not make it to this paper because the raw environment was not fair due to the ISP’s selective throttling and blocking. Exploring the reasons for why some protocols were blocked or throttled goes beyond the scope of this paper, but we had to look into these matters to create a platform for fair testing. 

Since OpenVPN was the most affected protocol among all, this matter had to be resolved. Options such as OpenVPN in TCP mode over Stunnel were considered and tested. The reason Stunnel was considered is because of its ability to wrap the TCP connection of the OpenVPN inside an SSL tunnel and successfully bypass the Deep Packet Inspection (DPI) used by ISPs to throttle or block connections. Hence, the plan to test all three VPN protocols over the above-described OpenVPN setup over Stunnel to provide a consistent environment was followed. In other words, it was planned to have every protocol X running on top of OpenVPN running inside Stunnel to provide a fair connection through a non-restrictive Internet node in New York; see Figure~\ref{fig:setup}.

Initially, the plan was to test these protocols on Raspberry Pi devices to mimic weak IoT devices, but because of the great complexity of this plan, this plan was changed. The reasons for complexity were added to because of OpenVPN’s great overhead in time consumption in its setup. The only working setup of OpenVPN on the available ISP, Stunnel, was an added complexity to implement across multiple Raspberry Pi devices. Instead, an eighth-generation Core i7 laptop with considerable capability was used to replace the Raspberry Pis. However, now we were missing the very reason for this research: seeing how Wireguard compares to traditional solutions on weaker devices. To counter this problem, a virtual machine was used on this device, and it was set to use only one of the eight logical cores of the processor. Ram was also limited to one gigabyte. This gave us a virtual machine that was even weaker than the Raspberry Pi 3 B+ models that were available. Hence, the challenge of simulating a weak IoT device was mostly overcome. Part of this was still an issue, and it was mentioned in Section II-E-1. Using a virtual machine on an Intel laptop, we still had a device capable of AES-NI instructions in its architecture, and this did not mimic the lack of these instructions on most IoT devices, giving a slight advantage to OpenVPN that would not exist in the typical IoT context.

Now, the issue of ISP’s blocking of certain protocols remained. The choice of a single machine for the setup of the virtual machine and testing helped this endeavor. A TCP OpenVPN connection was established on the main machine using Stunnel. The purpose was for the main machine to act as an uncensored gateway. All protocols would later be tested in the virtual machine through this gateway. Initially, Wireguard itself was considered to serve as the underlying connection for testing all protocols. This was because it had shown great performance and almost no bandwidth loss in pre-tests. However, this plan was changed. This was because Wireguard could not serve as both the main protocol to test and also the gateway. Hence, we chose the tried and tested OpenVPN in TCP mode over Stunnel to bypass ISP restrictions and try to create a fair environment for all the protocols on the virtual machine to be tested. 

\subsection{Final Experimental Setup}

\begin{figure}[h!]
\centering
\includegraphics[width=1.\columnwidth]{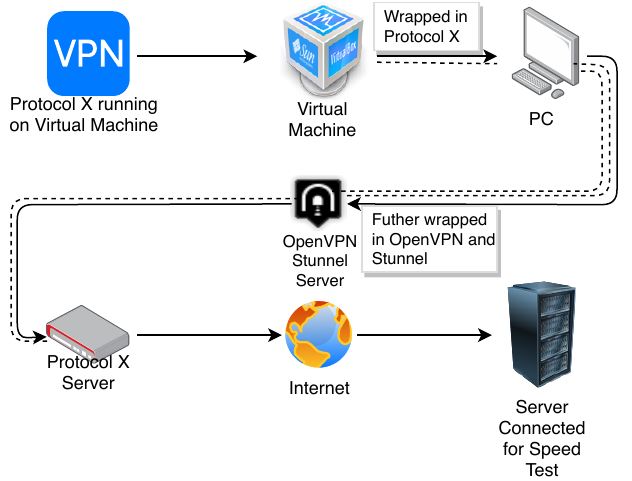}
\caption{Setup Used For Testing}
\label{fig:setup} 
\end{figure}

The final setup included the main machine connected to the Internet over the OpenVPN TCP connection over Stunnel, as shown in Figure \ref{fig:setup}. Over this open and unrestricted, albeit bottle-necked, gateway, the virtual machine then acted as a client to one of the three VPN servers for each of the protocols under examination. Only one VPN connection was activated on the virtual machine at a given time for testing.

For testing and data collection, Nperf \cite{b14} was used. Nperf is an online tool that provides access to thousands of gigabit and multi-gigabit servers for testing. It also runs several instances of a test at the same time to achieve the maximum performance possible on a link. In addition to throughput, Nperf runs twenty instances of a ping test and provides the following data:

\begin{itemize}
\item Average Download Speed
\item Peak Download Speed
\item Average Upload Speed
\item Peak Upload Speed
\item Latency (Ping)
\item Time Taken to Initialize the VPN Connection
\end{itemize}

All our VPN servers for OpenVPN, OpenConnect, and Wireguard had their servers in New York, and we chose a gigabit server in the same city that Nperf recommended. For each VPN, the test was repeated five times, and the data was collected.

\section{Results}
\label{sec:results}

\begin{figure}[tb]
\centering
\includegraphics[width=1.\columnwidth]{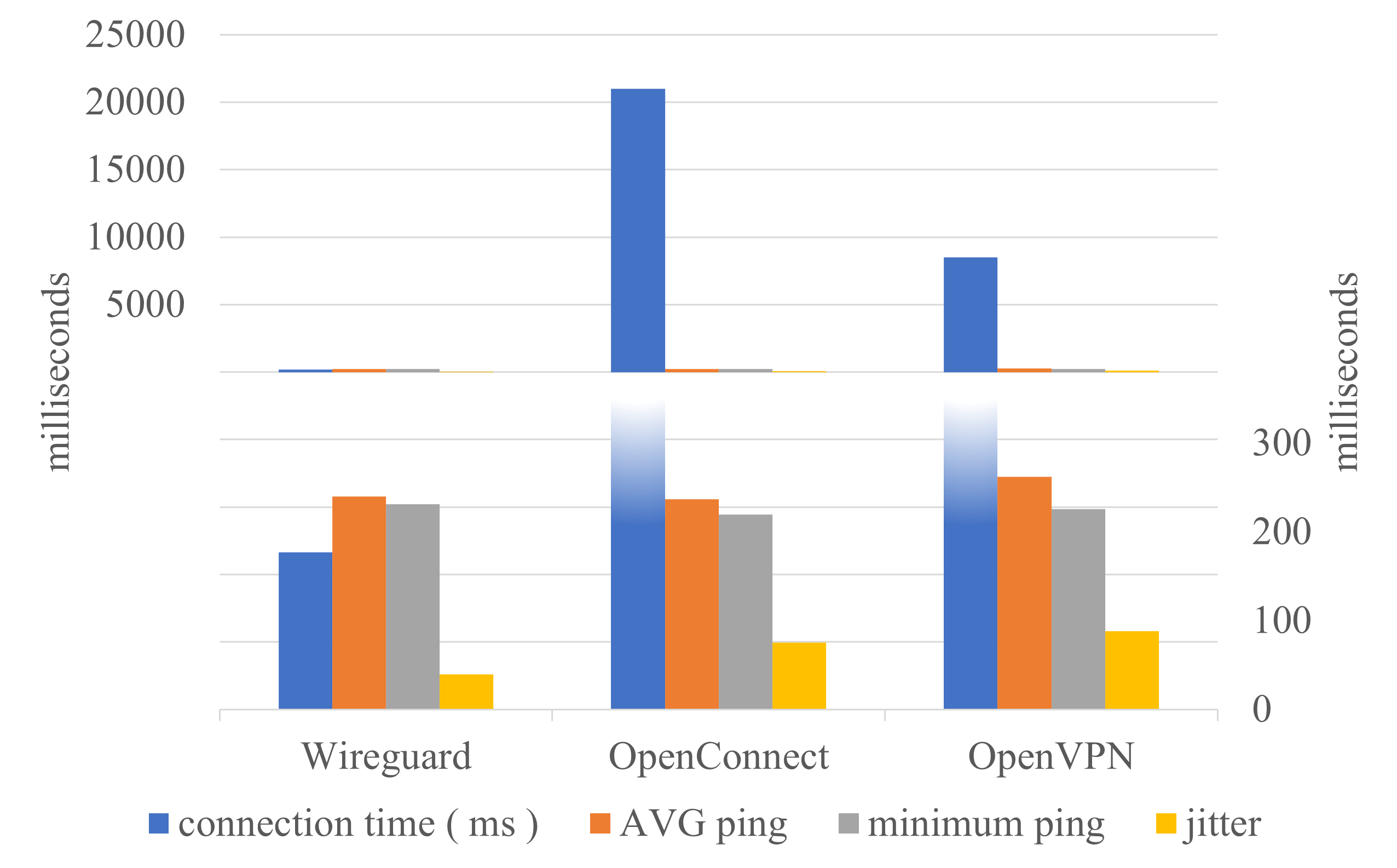}
\caption{Jitter (Variance in Ping)}
\label{fig:connectiontime}
\end{figure}

\begin{figure}[tb]
\centering
\includegraphics[width=1.\columnwidth]{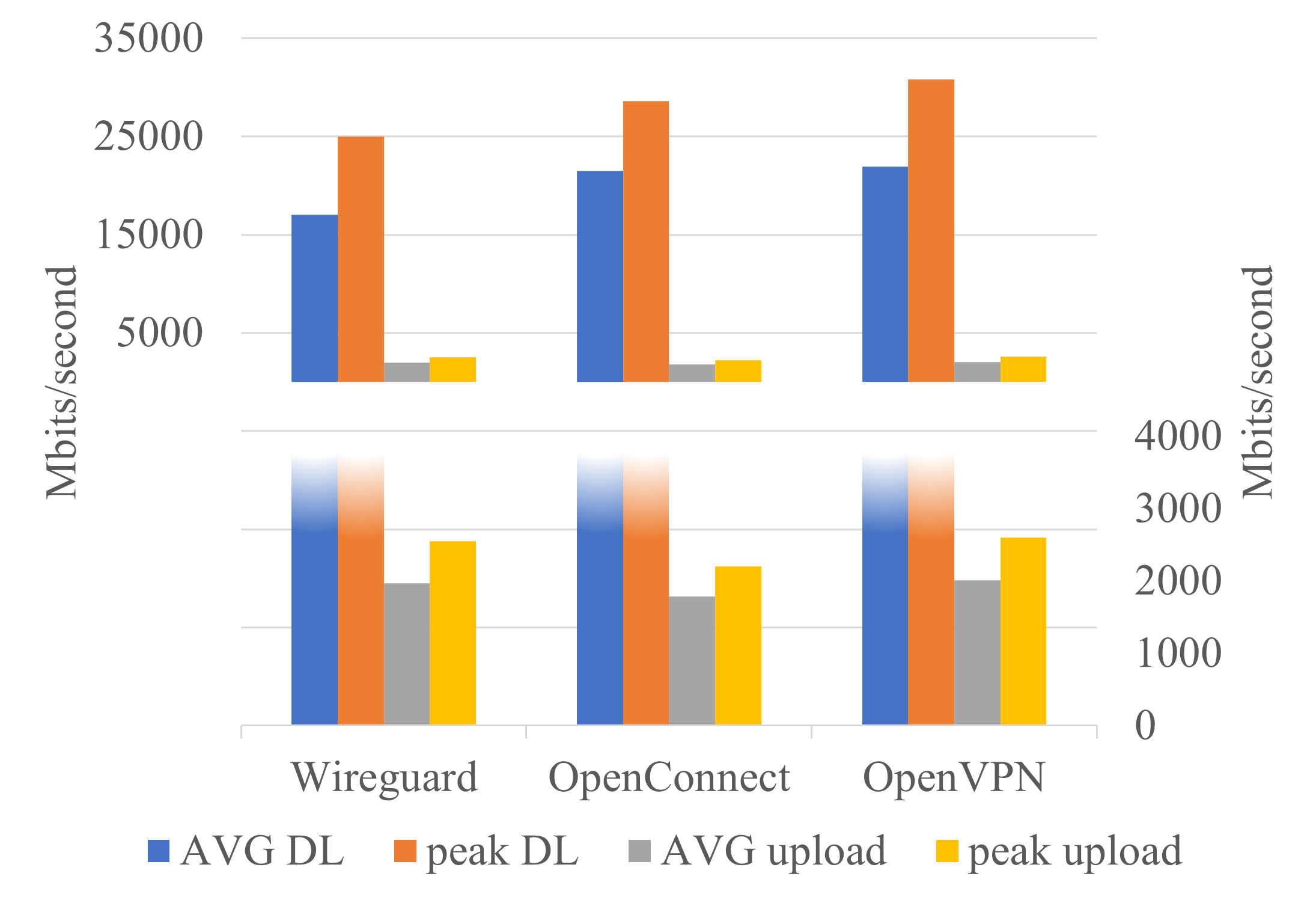}
\caption{Average Download Speed}
\label{fig:throughput} 
\end{figure}

\begin{figure}[tb]
\centering
\includegraphics[width=1.\columnwidth]{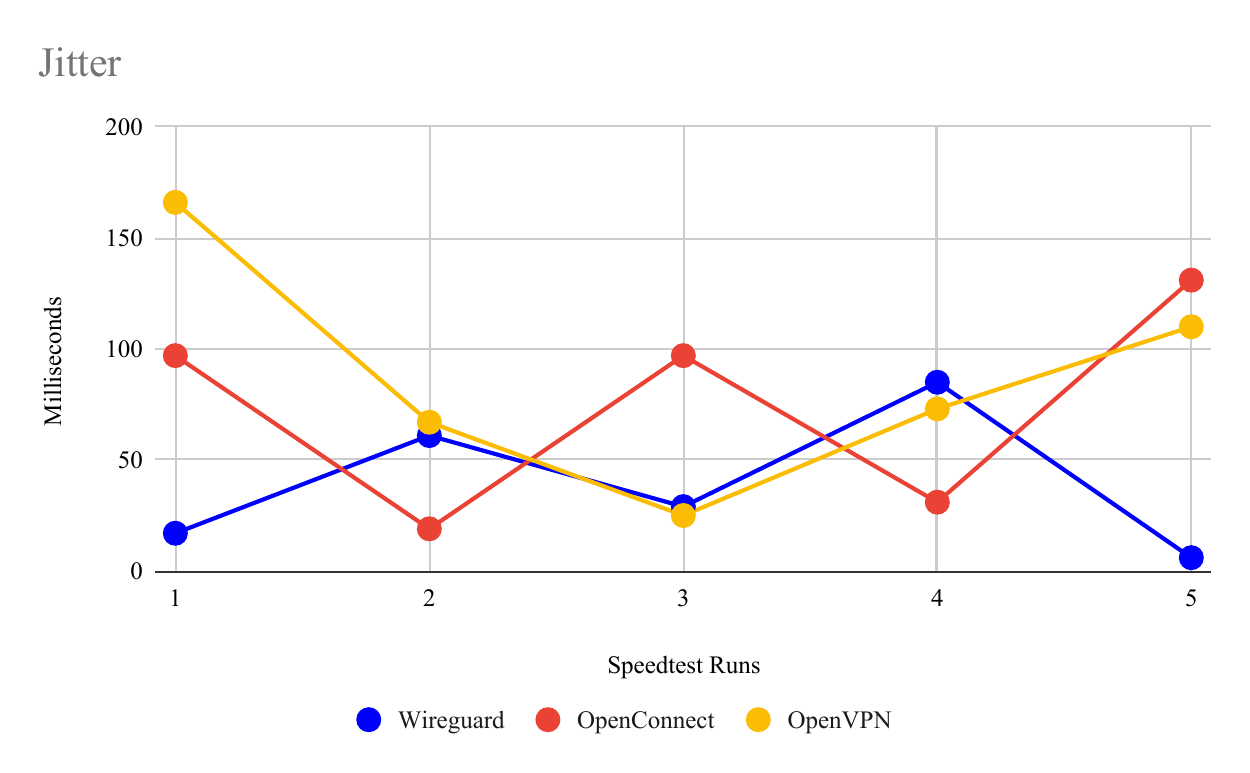}
\caption{Jitter (Variance in Ping)}
\label{fig:jitter}
\end{figure}

In our multifaceted evaluation, Wireguard performed with a clear advantage in connection setup time, jitter, and variance in jitter and had comparable performance in ping as shown in Figures~\ref{fig:connectiontime} and ~\ref{fig:jitter}. However, Wireguard was not able to challenge OpenVPN and OpenConnect in throughput, as shown in Figure~\ref{fig:throughput}.

%our three charts

%\begin{landscape}

\begin{table*}[bt]
\centering
\caption{Raw Data  for Each Instance of Five Tests per Protocol}
\resizebox{\linewidth}{!}{%
%\begin{tabular}{>{\hspace{0pt}}p{0.106\linewidth}>{\hspace{0pt}}p{0.138\linewidth}>{\hspace{0pt}}p{0.148\linewidth}>{\hspace{0pt}}p{0.11\linewidth}>{\hspace{0pt}}p{0.119\linewidth}>{\hspace{0pt}}p{0.117\linewidth}>{\hspace{0pt}}p{0.127\linewidth}>{\hspace{0pt}}p{0.062\linewidth}}
\begin{tabular}{lccccccr}
\toprule
 & AVG download & peak download & AVG upload & peak upload & AVG latency & peak latency & jitter \\ \midrule
 & 12173 & 15541 & 1735 & 2380 & 234 & 231 & 17 \\
 & 24327 & 32970 & 2078 & 2750 & 240 & 230 & 61 \\
Wireguard & 18354 & 27219 & 1863 & 2294 & 236 & 231 & 29 \\ 
 & 14628 & 23076 & 2062 & 2749 & 255 & 231 & 85 \\
 & 15529 & 26047 & 2101 & 2573 & 233 & 232 & 6 \\ \midrule
% &  &  &  &  &  &  &  \\
 & 13122 & 18972 & 1827 & 2233 & 236 & 219 & 97 \\
 & 19261 & 28995 & 1621 & 1868 & 228 & 219 & 19 \\
OpenConnect & 23102 & 30788 & 1685 & 2168 & 239 & 220 & 97 \\ 
 & 22614 & 29993 & 1940 & 2321 & 231 & 219 & 31 \\
 & 29356 & 34137 & 1862 & 2423 & 251 & 219 & 131 \\ \midrule
% &  &  &  &  &  &  &  \\
 & 21938 & 31884 & 2104 & 2811 & 361 & 233 & 166 \\
 & 21819 & 28091 & 1826 & 2636 & 234 & 221 & 67 \\
OpenVPN & 22880 & 35847 & 1979 & 2415 & 233 & 224 & 25 \\
 & 24998 & 31456 & 2085 & 2708 & 239 & 229 & 73 \\
 & 17922 & 26736 & 2042 & 2449 & 244 & 222 & 110 \\
\bottomrule
\end{tabular}
}
\label{tab:rawdata}
\end{table*}

\begin{table*}[bt]
\centering
\caption{Mean of All Tests per Protocol}
\resizebox{\linewidth}{!}{%
%\begin{tabular}{>{\hspace{0pt}}p{0.104\linewidth}>{\hspace{0pt}}p{0.192\linewidth}>{\hspace{0pt}}p{0.063\linewidth}>{\hspace{0pt}}p{0.073\linewidth}>{\hspace{0pt}}p{0.108\linewidth}>{\hspace{0pt}}p{0.119\linewidth}>{\hspace{0pt}}p{0.083\linewidth}>{\hspace{0pt}}p{0.121\linewidth}>{\hspace{0pt}}p{0.062\linewidth}} 
\begin{tabular}{lrrrrrrrr}
\toprule
 & connection time (ms) & AVG DL & peak DL & AVG upload & peak upload & AVG ping & minimum ping & jitter \\ \midrule
Wireguard & 177 & 17002.2 & 24970.6 & 1967.8 & 2549.2 & 239.6 & 231.0 & 39.6 \\
OpenConnect & 21000 & 21491.0 & 28577.0 & 1787.0 & 2202.6 & 237.0 & 219.2 & 75.0 \\
OpenVPN & 8500 & 21911.4 & 30802.8 & 2007.2 & 2603.8 & 262.2 & 225.8 & 88.2 \\
\bottomrule
\end{tabular}
}
\label{tab:meanrawdata}
\end{table*}

%\end{landscape}

The raw data is shown in Table~\ref{tab:rawdata} and the mean of the raw data is displayed in Table~\ref{tab:meanrawdata}. While Wireguard did follow the hypothesis for impressive connection setup and response time, these results are largely contrary to the results expected in this experiment in other aspects such as data throughput, especially in download (see Figure~\ref{fig:throughput}). Two reasons may explain this discrepancy in expectations and actual results:

\begin{enumerate}
\item MTU size. Wireguard's packet sizes are different from those of OpenVPN. This can affect throughput considerably because when packet sizes are not in sync, they are wrapped in multiple packets instead of just one, and this results in a slowdown of the connection. This and other factors inherent in the OpenVPN protocol could have given OpenVPN the edge here and hence, unfairly affected Wireguard's performance. The reason this is considered unfair is two OpenVPN protocols running over each other have similar properties in terms of datagram size and other properties, but for a Wireguard connection running on top of an OpenVPN base connection, the process is not that smooth, and sometimes the network has to re-transmit packets because of the inconsistent properties of Wireguard and the base OpenVPN connection used in testing all three protocols. 

\item AES-NI Instructions: Another advantage that we were looking for for Wireguard in the weak IoT devices was the lack of AES-NI instructions in many of these devices. However, in the machine we finally ran the tests on, AES-NI instructions were present in the system. This obstructed simulating an environment without these instructions where OpenVPN would not have a native advantage.  The reason this is considered an inherently unfair advantage in the setup is that in this paper, Wireguard is being presented as an efficient solution for weak IoT devices while the Intel device used in the test had OpenVPN-specific AES-NI instructions included in the hardware. This made the testing environment slightly less realistic with regards to IoT devices -- the main subject for the protocols in this paper.
\end{enumerate}

On the other hand, Wireguard did show more stability, as can be seen in Figures ~\ref{fig:connectiontime} and ~\ref{fig:jitter}. The jitter and variance in ping were the lowest for Wireguard consistently, and this can be seen in Table ~\ref{tab:rawdata}, too. Wireguard also showed good upload speeds (see Figure~\ref{fig:throughput}) and competed with OpenVPN and OpenConnect despite the disadvantages mentioned above. While Wireguard decisively proved to be the "quicker" protocol, it was affected in terms of throughput at higher speeds due to its magnified incompatibility with the underlying Stunnel over OpenVPN.

\section{Discussion and future work}

While the desired results based on the theoretical analysis could not be demonstrated in this paper, Wireguard's promise was not compromised due to the reasons discussed. Despite the reasons above, Wireguard still showed that it can compete with the other two mainstream VPN protocols reliably and even outperform them in certain aspects, such as:

\begin{enumerate}
    \item Simplicity
    \item Ping
    \item Jitter
    \item Connection Setup Time
\end{enumerate}

While Wireguard showed advantages in the above variables, it does not mean that Wireguard is superior in other aspects when used in weak IoT devices, for reasons discussed in Section ~\ref{sec:results}. 

For future work, it would be helpful to:

\begin{enumerate}
    \item Repeat the same data collection as is collected in this paper, but on an Internet infrastructure that is open and does not need an extra layer of another tunnel to provide an open connection to the protocols being tested.
    \item Conduct the tests on Raspberry Pi's or another type of similar non-Intel device that lacks AES-NI instructions.
    \item Collect CPU usage data when the tests are running to maybe show Wireguard's further advantage in lesser processing overhead.
    \item Demonstrate further advantages of Wireguard by efficiently securing blockchain networks in order to show Wireguard's compatibility.
\end{enumerate}

\section{Conclusion}

In this study, we evaluated Wireguard's potential as an efficient VPN solution tailored for resource-constrained IoT devices. The results demonstrate Wireguard's advantages in simplicity, connection speed, and stability compared to OpenVPN and IPsec. Although experimental limitations prevented fully simulating an IoT environment and showing Wireguard's full potential, Wireguard still showed promise in aspects like low overhead and jitter. With further testing under ideal conditions, Wireguard may emerge as a lightweight yet robust IoT security solution. Its efficiency could enable widespread VPN adoption to harden vulnerable IoT devices against attacks.

\end{document}